\documentstyle[11pt,newpasp,twoside,epsf]{article}
\def\edcomment#1{\iffalse\marginpar{\raggedright\sl#1\/}\else\relax\fi}
\reversemarginpar

\begin{document}
\title{Conference Summary}
\author{Richard S Ellis}
\affil{Astronomy Department, California Institute of Technology,
Pasadena CA 91125, USA}
%\author{Ima Co-Author}
%\affil{The Name of My Institution, The Full Address of My Institution}

%\begin{abstract}

%\end{abstract}

\section{Introduction}

Clusters of galaxies are of wide interest in the astronomical
community both as laboratories where baryons and dark matter can
be conveniently studied and, through their statistical properties
at various look-backtimes, as tracers of cosmic structure and
evolution. The Local Organising Committee wisely narrowed the
topics of this meeting to four broad themes and so I will organise
my summary remarks under these headings.

First I think I should address the plea "Why bother to study
clusters?", raised provocatively by Colin Norman at one stage
during the proceedings\footnote{Fortunately, I have forgotten to
which speaker he made this remark!}. We have learnt enough about
clusters and the growth of structure to know that two of the
traditional motivations require much more careful consideration.

For years on time allocation committees, I read observers repeat a
{\em mantra} in the first sentence of their proposals: {\em
``Clusters of galaxies represent the largest bound structures in
the Universe.."} in justification of their role as tracers of
large scale structure. A number of speakers at this meeting
reminded us that clusters no longer uniquely occupy this role. We
have other tracers of large scale structure and, moreover, it
seems we need to be much clearer of what exactly we mean by a
cluster before we can convincingly use them as cosmological
probes.

A second, traditional, motive for studying clusters has been
observational convenience, e.g. in studying constituent
populations such as galaxies at various look-back times. At first
sight, it's an attractive proposition for an observer to study a
few rich clusters at various redshifts each containing hundreds of
accessibly-luminous galaxies and then to ``join the results" in a
timeline to make some evolutionary claim. But, at most redshifts
of interest, we can expect to find a wide range of overdensities
represented (groups, merging systems, virialised clusters). Put
simply, rich cluster A at $z$=1.2 is unlikely to evolve into rich
cluster B at $z$=0.5 and neither may necessarily become a
present-day Coma.

Both these worries indicate a high degree of rigour is needed in
using clusters. We need very large samples spanning ranges of mass
at any redshift of interest and perhaps it would be foolish to
adopt one single selection criterion for their study. A similar
``panchromatic" theme has emerged over the past decade in
understanding how to address the question of evolution using
samples of galaxies.\footnote{Witness the controversy surrounding
the use of UV/optical and far-infrared probes of the cosmic star
formation history. No single technique wins: several are
required.}

The motivation for studying clusters that emerged at this meeting
focused broadly along the following themes:

\begin{itemize}

\item{} Testing gravitational instability by measuring the number
density of massive clusters at high redshift, viz.
$\Phi(M_{cluster},z)$.

\item{} Breaking degeneracies in estimates of the the cosmological
parameters by examining the local population of clusters as a
probe of the mean mass density and the normalisation of the mass
power spectrum, $\Omega_M$ and $\sigma_8$.

\item{} Verifying hierarchical structure and the nature of the dark matter
and keeping those theorists in check who predict a universal mass
profile with central cusps, $\rho_M(r) \propto r^{-1.5}$.

\item{} Determining the origin of the heating of the intracluster
medium, its enrichment history and examining whether
non-gravitational processes are involved.

\item{} Examining the history of spheroidal galaxies and role of
the environment, for example in understanding the origin of the
morphology-density relation and the destiny of the infalling
component of gas-rich field galaxies.

\end{itemize}

Let me make a disclaimer in what follows. I am not an expert in
any of the areas I was asked to summarise and so I submit these
concluding remarks only as someone who tried to listen carefully
to most talks, dutifully avoiding the lunchtime mountain hikes to
make sense of what I heard\footnote{I apologise if I missed some
talks but I tried to secure powerpoint files or transparencies for
most of those.}. With $\simeq$70 talks and $\simeq$80 posters,
inevitably I have had to be very selective in discussing results.

If one could summarise the meeting in two paragraphs, I would say
the following:

\begin{itemize}

\item{} There is an explosion of terrific data, from large
surveys (X-ray and ground-based) which offer qualitatively new
ways in which we find and do statistical studies of clusters, and
also in {\em resolved} data within clusters which opens up new
opportunities for understanding the detailed astrophysics of dense
environments.

\item{} The subject is moving from exploratory surveying
to detailed astrophysics. As part of this ``growing up" there is a
need to admit defeat on some of the old methods and to embrace new
ones, particularly on the statistical questions. For many years
cluster workers had something of a monopoly in the study of early
galaxies and large scale structure, but it is time to take
advantage of other datasets being delivered and view cluster
astrophysics as only one part of a larger body of information.

\end{itemize}

\section{Searching for Distant Clusters}

During the first part of our meeting we listened to progress made
in finding clusters, mostly (but not exclusively) at high
redshift. We heard about a bewildering number of deep surveys
(MACS, REFLEX, WARPS, LCDCS, EIS, RDCS to name but a few..Figure
1) each requiring significant follow-up with the world's most
powerful facilities. Excellent reviews of this active field have
been given by Borgani (2001) and Borgani \& Guzzo (2001).

\begin{figure}
\plotone{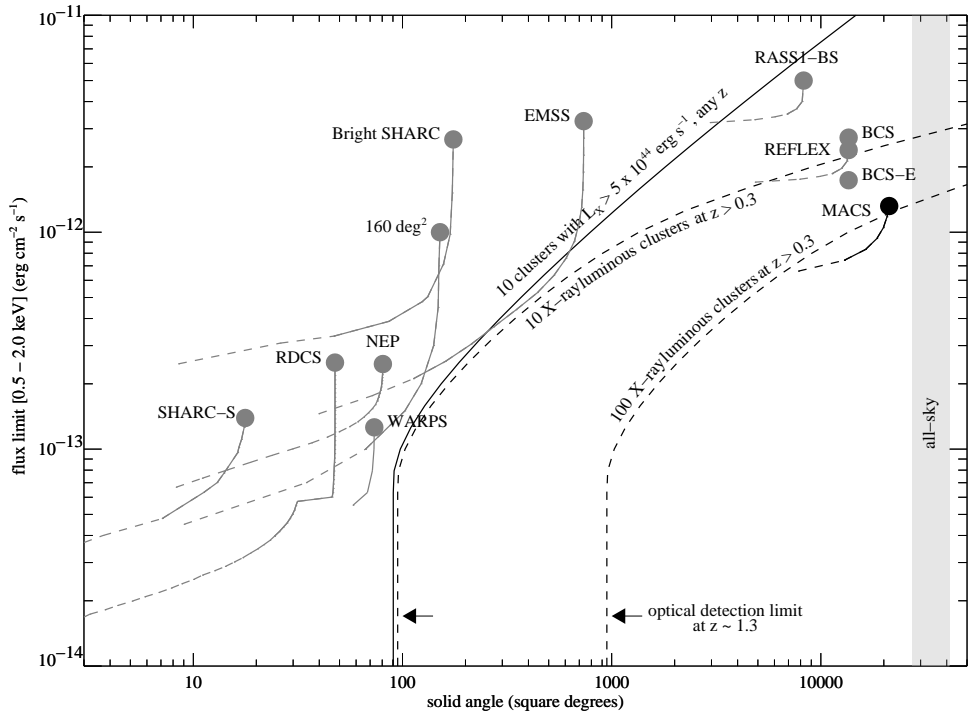} \caption{The impressive effort to track
down the evolution of X-ray luminous clusters to high redshifts:
survey parameters taken from the compilation of Harald Ebeling.}
\end{figure}

The motivation is a sound one: testing gravitational instability
in the context of a known cosmological model, or alternatively, by
assuming a structure formation model such as CDM, constraining the
cosmological parameters. The fraction of collapsed large massive
structures ($M\simeq10^{14}M_{\odot}$) at, say, $z\simeq$1 for a
given cosmology is one of the most robust predictions of the dark
matter models (Figure 2). The necessary ingredients are a large
well-defined sample and robust estimates of the cluster masses.

\begin{figure}
\plotfiddle{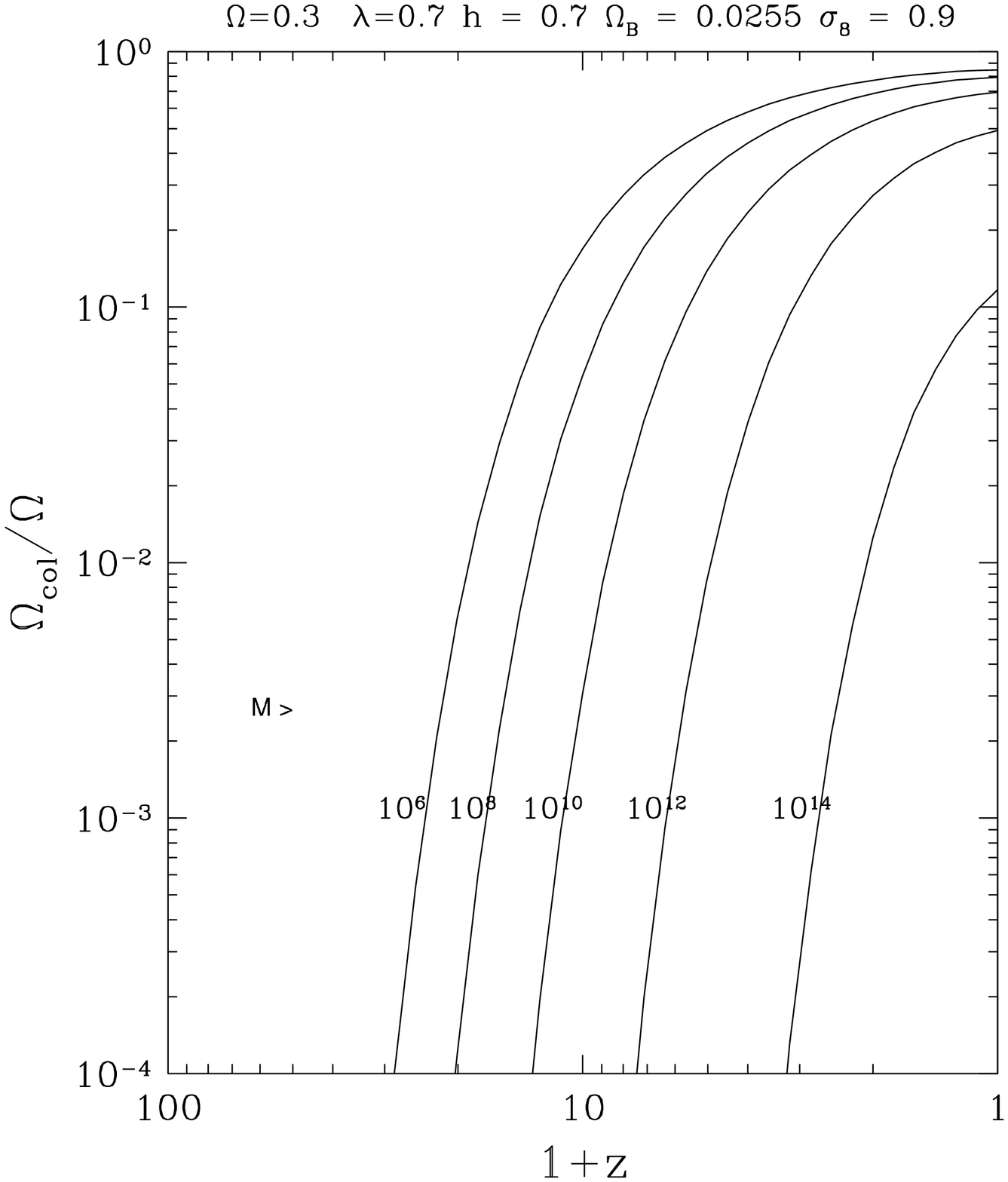}{1truein}{0}{25}{25}{-170}{-120}
\plotfiddle{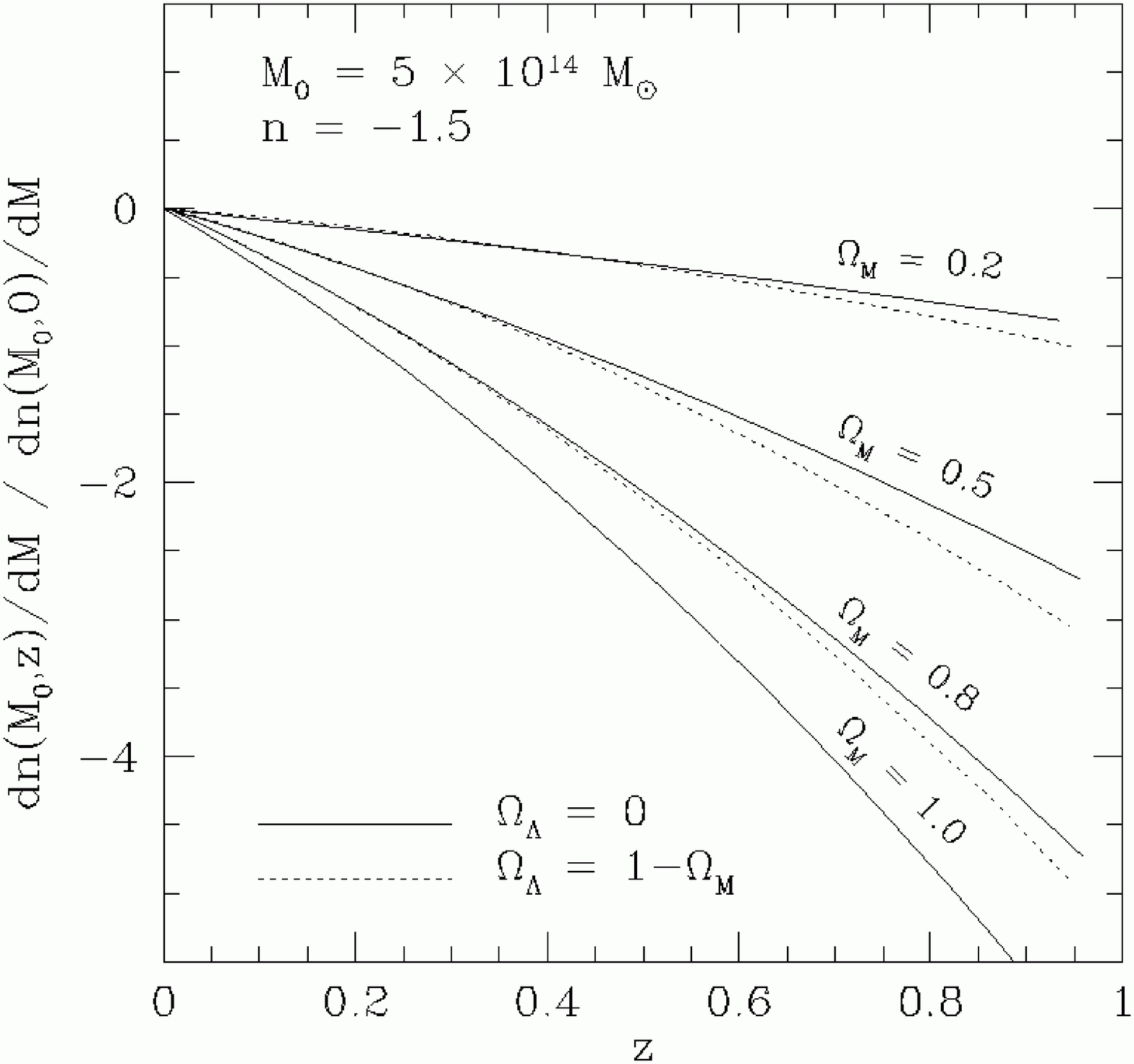}{1truein}{0}{20}{20}{00}{0}
%\plottwo{sesto_fig2a.eps}{sesto_fig2b.eps}
\caption{(Left) The fraction of gravitationally collapsed objects
(measured in terms of $\Omega_M$) with mass greater than a given
value (in solar mass units) as a function of redshift, $1+z$, is
one of the simplest and most robust predictions of CDM theory
given a set of cosmological parameters ($\Lambda$CDM here, after
Fukugita 2000). (Right) The wanted evolutionary signal, a decline
in the number of massive clusters with redshift, as a function of
cosmological model (after Voit 2000).}
\end{figure}

An additional motivation for locating high $z$ clusters, is to
undertake evolutionary studies. For galaxies, this is less
appealing now we can find statistically-complete field samples at
similar redshifts by independent photometric means (Adelberger
2000, Daddy et al 2001, McCarthy et al 2001), many of which are
extensive enough for locating clustered objects in a
self-consistent way. Marc Postman stressed the dangers here by
contrasting differences found between conclusions drawn from
galaxies studied in the X-ray selected CNOC and optically-selected
Morphs samples. If clusters are found by a manner which relates to
the properties of their constituent galaxies (e.g. by only
searching for ones with prominent red sequences as discussed by
Mike Gladders, or finding ones associated with powerful radio
galaxies as reviewed by Philip Best), one can get different views
of what is going on in dense regions.

The quest for high $z$ clusters has a chequered history. Most of
the early work was motivated by a quest for the deceleration
parameter based on brightest cluster galaxies (Gunn \& Oke 1975,
Kristian et al 1976). Denis Zaritsky gave us a new twist to this
story with his evidence for varying accretion onto brightest
cluster galaxies in the Las Campanas Distant Cluster Survey (see
also Arag\'on-Salamanca et al 1998). Claims for number evolution
in cluster counts with redshift have also swung to and fro over
the years. Until recently only a handful of optical or X-ray
selected clusters were available\footnote{On more than one
occasion, challenges to a standard paradigm have been made on the
basis of the existence of a single high $z$, assumed massive,
cluster.}. The most exciting aspect about this meeting is the
enormous advance in the numbers of distant clusters.

So how best to find distant clusters? Marc Postman nicely
summarised the various search techniques, their advantages and
biases. The most useful techniques are those that minimise
projection effects. Faint X-ray surveying in the 0.1-10 keV range
is cerainly an expensive method in telescope time (but good things
don't come cheap): the background is low, we expect $L_x \propto
n_e^2$ and thus, so long as massive clusters have hot intracluster
media, this is the method of choice. Kathy Romer emphasized how,
even with the higher EPIC background, XMM promises to extend the
depth of the ROSAT surveys to mean redshifts well beyond
$z\simeq$1.

Whilst we have yet to reap the benefits of SZ effect as a survey
tool, James Bartlett showed us that this method has the great
advantage of being distance independent (modulo evolution) with a
signal $\propto n_e$. Instruments such as Planck Surveyor, and the
proposed Arcminute Microkelvin Imager which R\"udiger Kneissl
described, should also readily probe beyond $z\simeq$1.

In a similar category to the SZ effect is weak lensing. As this
technique wasn't explicitly reviewed at the meeting, I reproduce
the nice discovery of a $z$=0.28 cluster by Wittman et al (2001)
in Figure 3. Although lensing offers the most direct route to the
total mass, it still suffers from projection effects, the effect
of the mass sheet degeneracy\footnote{Mass estimates based on weak
shear of a background population are insensitive to an additional
unclustered component of dark matter.} and, as a high density of
background galaxies is needed, the technique will most likely be
limited to the study of clusters with 0.1$<z<$0.8.

\begin{figure}
\plotfiddle{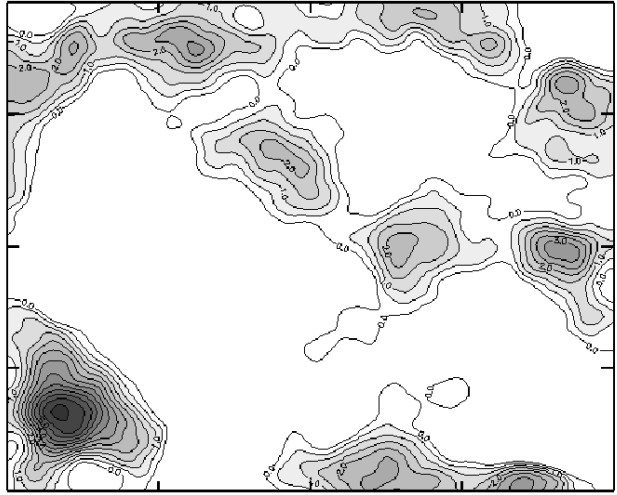}{2.5truein}{0}{35}{35}{-100}{0}
\caption{The projected distribution of dark matter in a 40 arcmin
field determined from the shear of faint galaxies by Wittman et al
(2001). The tangential shear as a function of the photometric
redshift of the background galaxies places the 4.5$\sigma$
concentration in the lower left corner at a redshift of
$z\simeq0.3\pm0.08$. Spectroscopy of the cluster in this region
yields $z_c$=0.28 with a dispersion $\sigma_c$=615 km s$^{-1}$.}
\end{figure}

Concerning the optical searches, Marc showed us, from the analysis
of Goto et al (in preparation) how, even locally within the SDSS,
changing the search algorithms can deliver very different samples.
This is worrying to me as I imagine this kind of uncertainty can
only be worse in faint surveys. While smart ideas such as
``matched filters" designed to minimise projection and maximise
contrast, are of course to be welcomed there is a natural concern
that, like Maximum Entropy restoration techniques, such methods
work reliably only when we have a clear idea of what we are
searching for. Do we?

The Red Cluster Sequence method described by Mike Gladders seems
the most promising of the optical methods. Indeed the authors
claim it to be ``comparable or superior to X-ray methods"
(Gladders \& Yee 2000) - a bold assertion! Certainly panoramic
detectors sensitive in $R$ and $z'$ make this a much more
efficient way to locate red colour-magnitude sequences to
$z\simeq$1 and, at CFHT, there is the prospect of coupling lower
$z$ detections with weak lensing constraints to directly get
masses. Simulations suggest this search technique should be
complete to $z\simeq$1, even allowing for some dispersion and
bluing in the color-magnitude relation. Unless there is a
(perverse) population of massive clusters devoid of a uniform
population of spheroidals, the technique seems a sound one.

But the next thorny issue is how to get a {\em reliable mass}, the
second important requirement to test Figure 2 and one worth
thinking about before embarking on any time-consuming survey! The
mass of a distant cluster has traditionally been inferred from the
X-ray luminosity (assuming a $L_x-T_x$ relation), an optical
richness (very crude) or a velocity dispersion based on a few
members. Clearly none of these is really adequate. This is, I
think, where the X-ray surveys win hands down even if ``smart''
algorithms rescue the optical searches. We can fully expect, with
adequate investment, X-ray temperatures for large numbers of
distant clusters in the coming years. Brian Holden showed new
Chandra data that indicates little evolution in the $L_x-T_x$
relation to $z\simeq$0.8. Whilst assumptions are necessary to link
$T_x$ and virial mass (Voit 2000), the prospects look good,
particularly with cross-checks from the SZ effect. Lauro Grego
showed us the promise of deriving masses independently from her
interferometric SZ techniques.

The major problem in my opinion is that the optical searchers have
no real route to cluster mass (except by recourse to independent
X-ray techniques or perhaps weak lensing over a limited $z$
range); projection also remains an issue. Multiple systems, often
separated by only modest velocity differences, would be very hard
to detect without a huge spectroscopic investment. A good example
in showing the pitfalls is the cluster surrounding the $z$=1.206
galaxy 3C324 (Smail \& Dickinson 1995, Figure 4) now known from
spectroscopic evidence to consist of two components separated by
only $\delta z\simeq$0.06. Although radio-selected, not only would
its richness have been overestimated without careful spectroscopy,
it seems unlikely the Red Sequence method {\em applied to a
similar system at any redshift} would be able to
photometrically-separate the two systems.

\begin{figure}
%\plottwo{sesto_fig4a.eps}{sesto_fig4b.eps}
\plotfiddle{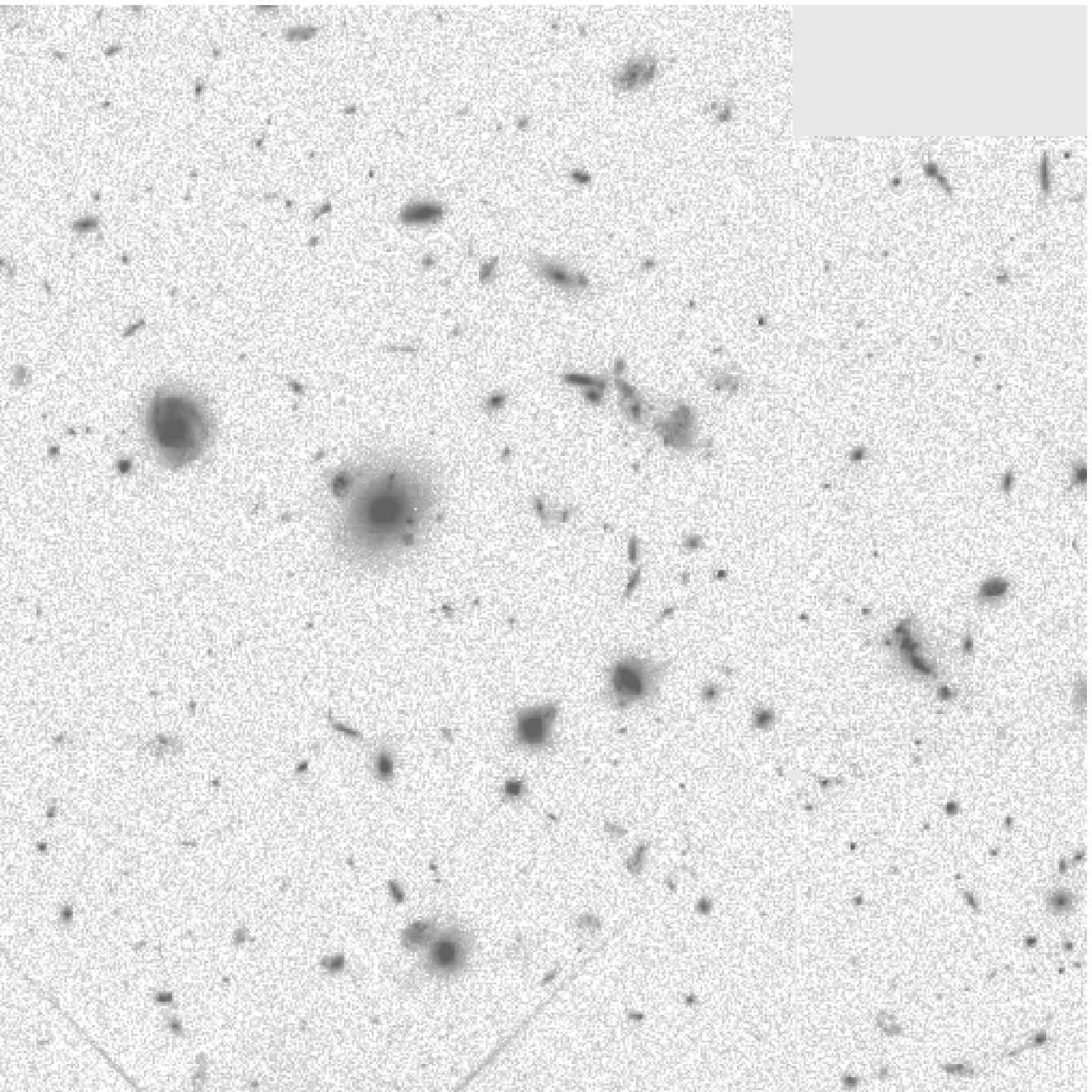}{1truein}{0}{25}{25}{-170}{-80}
\plotfiddle{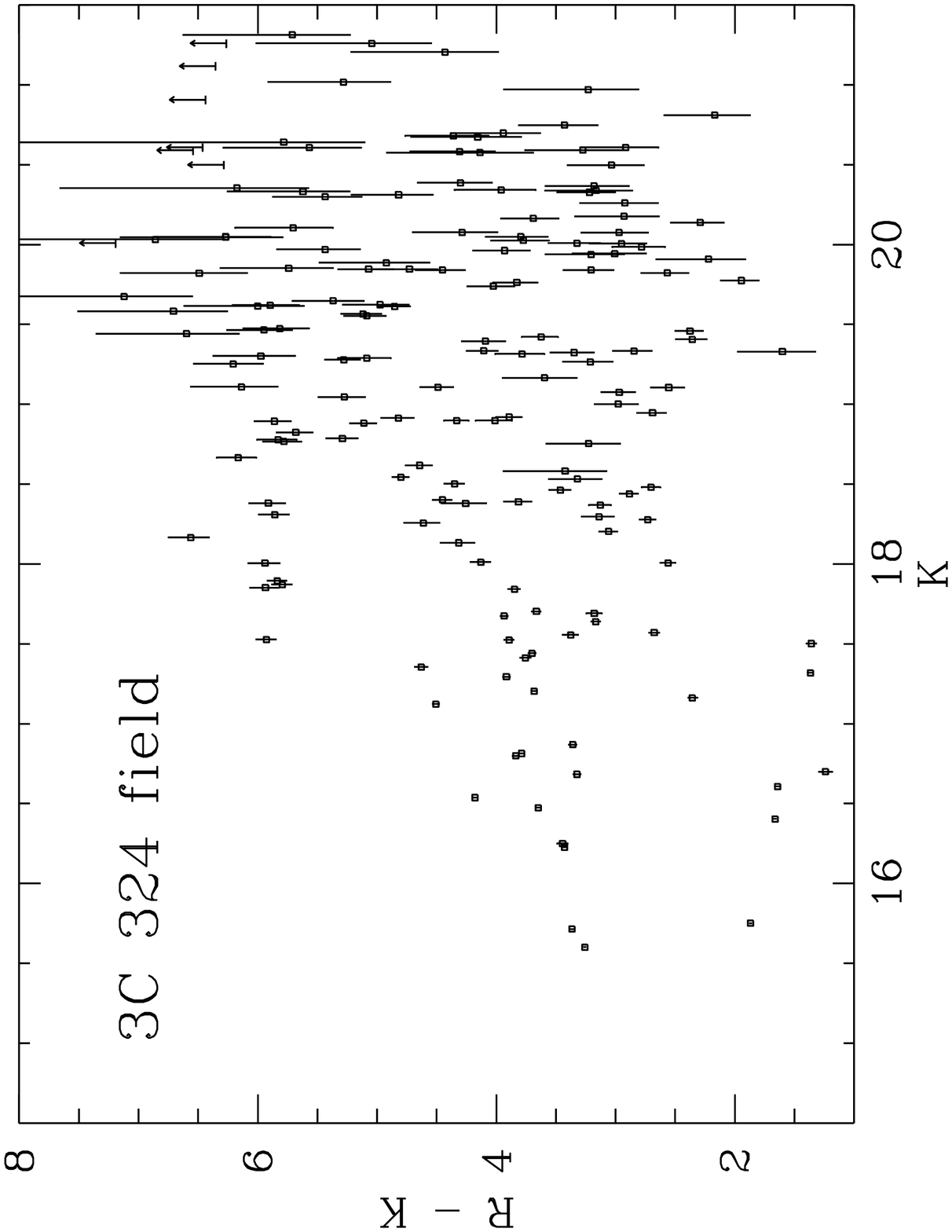}{1truein}{-90}{25}{25}{-20}{140}
\caption{3C324: an example of a complex system whose significance
would most likely be overestimated in any optical or
infrared-based search. The colour-magnitude relation (right) masks
the presence of two separate systems (with $R-K\sim$6) revealed
only with spectroscopy: one at $z$=1.206 containing the radio
galaxy and a second, separate system at $z$=1.115 (Courtesy of
Mark Dickinson)}
\end{figure}

In the case of the local X-ray cluster data essential as a basis
for all high $z$ comparisons, Luigi Guzzo and Hans B\"ohringer
demonstrated the remarkably precise cluster luminosity function
obtained from the REFLEX survey. Few would disagree that this is a
major achievement in the subject not just in statistical terms but
also because of the considerable care taken to ensure homogeneity
in the survey.

Turning then to the cosmological results and necessarily being
somewhat selective, Harald Ebeling (MACS) and Piero Rosati (RDCS)
discussed a fairly modest decline in the abundance of massive
clusters to $z\simeq$1 (Figure 5); this is a fairly convincing
conclusion for $z<$0.5 but beyond I suspect there is still some
room for manoever. Qualitatively, this is similar to the
4.7$\sigma$ decline claimed by Gioia (2001) in her NEP sample (not
discussed at this meeting). Stefano Borgani has modelled the
decline with redshift in the RDCS-3 sample with available
$L_x-T_x$ data and claims the data is consistent with popular
choices of $\Omega_M$ and $\sigma_8$ (see $\S$3).

\begin{figure}
\plotone{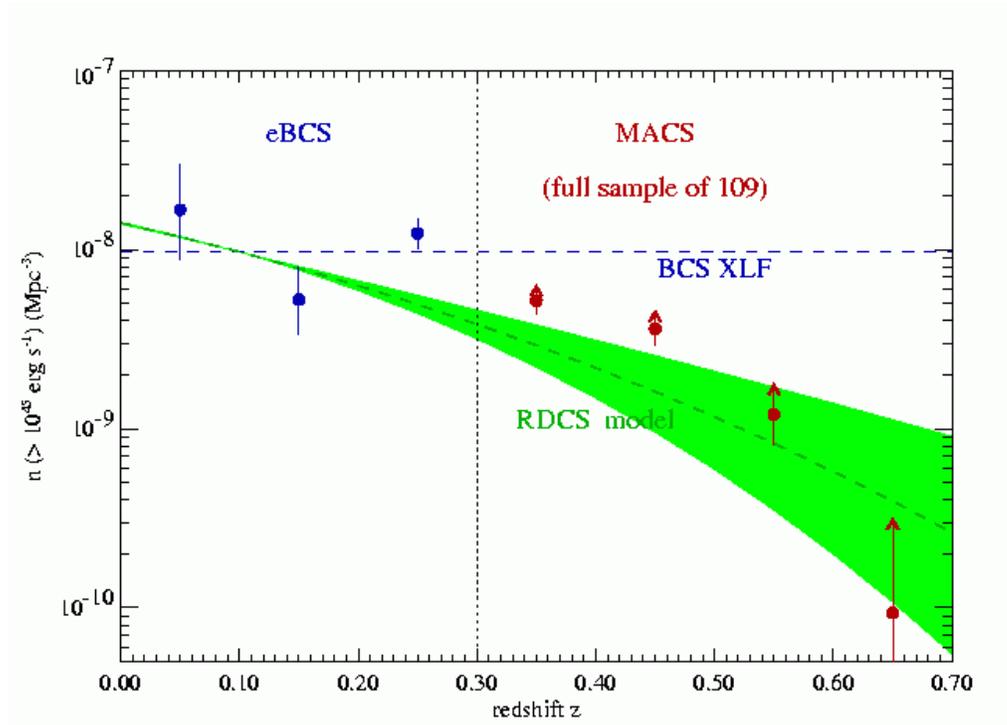}
\caption{Evolution in the comoving space density of the most X-ray
luminous clusters ($L_x>10^{45}$ cgs) from the MACS survey
(Ebeling), a result which indicates a slightly more modest decline
in number to $z\simeq$0.5 than that adopted by the RDCS team
discussed at the meeting. Broadly comparable results have been
obtained for the NEP sample (Gioia 2001). }
\end{figure}

Can we say there is ``concordance" between the observers? Not yet
but good progress is being made. Most of any agreement refers to
evolution in luminous, presumed massive, clusters. A number of
niggly issues remain including the effects of incompleteness,
cosmic variance for the smaller field surveys, AGN/cooling flow
contamination, and how to compare surveys with minimal $L_x,z$
overlap. For example, at the lower luminosities probed by the
WARPS survey, the situation is quite unclear with some arguing for
no evolution at all. Much of the dispersion in the inferred
evolution arises from uncertainties in the number of {\em local}
massive clusters against which comparisons are needed. WARPS and
RDCS appear to disagree on the ``local" abundance to an extent
that seriously affects their respective analyses. However, the
prospects for extending these tests to higher redshift and
clarifying the temperatures with Chandra appear very promising,
even before the new XMM surveys are underway.

As an outsider to the subject of distant X-ray cluster surveys, I
was struck in both Harald and Marc Postman's reviews of the large
number of competing surveys underway. If the labour involved in
constructing these surveys was not enough, remember that the time
required to fully exploit them is even larger. For example, the
construction of Ebeling's ROSAT-based MACS sample of 840 clusters
involved two colour CCD imaging and spectroscopic verification of
a significant sample. To exploit all 109 clusters beyond $z>$0.3
with multi-slit spectrographs, a weak lensing deep imaging study,
Sunyaev-Zel`dovich (SZ) detections and other applications would
require almost dedicated access to a huge range of facilities (a
tall order even by Harald's standards!). When one realizes this is
just {\it one} of the many ambitious surveys we heard about with
more in the pipeline, I worry has the planning of all these
surveys seriously taken into account the necessary investment for
a full exploitation?

\section{Mapping Large Scale Structure with Clusters}

Bob Nichol gave us a very balanced review of the changing role of
clusters as new probes of large scale structure became available.
He also posed some controversial questions, e.g: ``in the era of
galaxy and lensing surveys, who needs clusters?" He also stressed
many of the complications that arise from the frequent merging of
clusters on our assumption of relaxed systems in virial
equilibrium.

I think it fair to say the traditional role of using clusters
selected in various way (optically from the Abell/ACO catalogues
or from X-rays) as the most efficient way to get to the mass power
spectrum $P(k)$ on large scales is being overtaken by the large
redshift surveys. In the era of 2dF and SDSS, huge volumes
populated by over 100,000 galaxies are publically
available\footnote{See http://www.mso.anu.edu.au/2dFGRS/ \&
http://archive.stsci.edu/sdss/edr\_main.html.} and delineate the
power spectrum more reliably than the earlier Abell/ACO cluster
samples whose integrity is still being debated. In the case of the
X-ray cluster samples, Luigi Guzzo's analysis of the REFLEX survey
is comparable in importance to 2dF and SDSS at the time of
writing. A particularly impressive achievement in that sample is
the detection of infall in the $\xi_v(\sigma,\pi)$ plane
convincingly demonstrating the quality of the sample. However, the
redshift surveys are continuing apace and from them one can create
far more reliable spectroscopically-based cluster samples. Of
particular interest is the volume to $z\simeq$0.55 probed by SDSS
Luminous Red Galaxies survey. The message is clear: wholesale
galaxy mapping has arrived..use the data!

Bob drew attention to possible ``baryon wiggles" in both the power
spectrum of Abell/ACO clusters seen by Miller et al (2001) and
referred to one tentatively in the 2dF data (Percival et al 2001).
It must be remembered that, in such plots, the redshift-space
power spectrum is significantly affected by aliasing introduced by
the window function of the survey volume. Without careful
simulation of this effect it is very hard to be sure whether bumps
are artefacts or genuine baryon oscillations. Neither the REFLEX
not the 2dF surveys claim to have detected these features. Indeed,
the 2dF team estimates unless the baryon density is much higher
than expected, they will remain undetected in the completed SDSS
survey.

Elena Pierpaoli described a second aspect of local cluster
statistics, namely constraining the {\em amplitude} of the mass
power spectrum. The abundance of clusters of known mass provides a
joint constraint on the variance, $\sigma_8$ on 8$h^{-1}$ Mpc
scales and the mean mass density, $\Omega_M$ in a form dependent
on $\sigma_8\,\Omega_M^{0.5}$. By adding X-ray temperature
measurements and improving the mass-$T_x$ connection, an improved
constraints has been determined. In the past year, the same
constraint has also been probed in a completely independent way
via weak lensing studies of randomly-chosen fields (so called
``cosmic shear'' surveys) (eg. van Waerbeke et al 2001, Bacon et
al 2000, 2001 in prep.). Reassuringly both methods get fairly
similar answers with $\sigma_8\,\Omega_M^{0.6}$=0.40-0.50 although
it is important to remember that cosmic variance and the redshift
distribution of the background faint galaxies may affect
uncertainties in the lensing estimates\footnote{Since the
conference, estimates about 20\% lower for the combination of
$\sigma_8$\ and $\Omega_M$ have been published, both from cluster
studies and large scale structure results. Lahav et al
(astro-ph/0112162 gives a good summary)}

The weak lensing results are independent of any assumed cluster
population and do not rely on Gaussian fluctuations in the mass
spectrum. Accordingly, comparisons of lensing and cluster-based
methods are an important way to verify the assumptions we may take
for granted, not only in the cosmological framework but also in
cluster physics.

\section{Physical Processes Within Clusters}

We had an interesting session on the internal mass distributions
within cluster where Andrea Biviano gave a very comprehensive
review of years of effort invested in estimating orbital
anisotropies and mass distributions in clusters. The number of
clusters which have been comprehensively surveyed
spectroscopically is increasing rapidly and the new field surveys
(SDSS, 2dF) will generate lots more. Cluster galaxy dynamics will
remain a critical tool in determining the orbital anistropy of
cluster galaxies of different types. As Andrea showed, there is
convincing evidence that spheroidal galaxies retain an isotropic
velocity field, consistent with their long-standing membership,
whereas late type systems show a detectable radial anisotropy
reflecting their continuous infall.

The peripheries of clusters at moderate redshifts are largely
unstudied regions important in linking cluster-based environmental
evolution to that in field galaxies. Tommaso Treu and Taddy Kodama
described complementary dynamical/HST and photometric imaging
programs to address this question. Kodama finds a potentially
important result whereby the colours of field galaxies undergo a
sharp transition from blue-to-red at some galaxian surface
density, presumably reflecting some process associated with their
arrival at the cluster. Treu is embarking on a longer term project
to understand the mechanistic details of infall in a single
well-studied cluster. Both routes are important ones to pursue.

Andrea was cautious about deriving mass profiles from galaxy
velocities. This has been a long standing problem in individual
clusters, even Coma, because of the inevitable degeneracies
arising from unknown orbital anisotropies as a function of radius.
If a functional form is adopted for the mass profile (e.g. NFW),
solutions can be found but the present motivation is surely to
determine these forms directly, as is possible in a more
competitive way with X-ray data in the inner regions and weak
lensing to the periphery. Steve Allen showed how much progress has
been made in reconciling strong lensing and X-ray based mass
estimates following improved Chandra data on 7 clusters and found
no significant deviations from the universal mass profile seen in
numerical simulations.

In an effort to overcome some limitations, it has become common
practice to co-add data from many clusters in the hope of
improving signal to noise and erase asymmetries. Peter Katgert and
Roland van der Marel showed us results from coadded samples based
on the ENACS and CNOC surveys respectively, and concluded mass
traces light to a reasonable approximation. However, is a
composite cluster really a physical entity? A number of critics
were unconvinced because of the dangerous effect one or two
``complex" systems might have on the final conclusion. Even with a
substantial number of redshifts, it seems one has to be cautious
in interpreting dynamical data in terms of a simple gravitational
potential.

A salutory lesson can be learnt from the comprehensive survey of
the regular cluster Cl0024+16 (see Treu et al, Metevier et al,
these proceedings) by Czoske et al (2001). For many years this was
regarded as a classic regular virialised system at $z$=0.4 and its
mass profile even formed the basis for promoting self-interacting
dark matter as discussed by Oleg Gnedin. However, when 650
redshifts were gathered (of which 295 are members), Czoske et al
resolved a foreground system (whose mass was initially thought to
be quite modest but now is claimed to be significant) undergoing
an end-on merger with the main body of the cluster (Figure 6).
This discovery may explain its low X-ray luminosity c.f. the
original high velocity dispersion. With less than 100 member
velocities, Cl0024+16 would surely have been considered a
representative cluster for coaddition but now we know it is a more
complex beast. Bootstrap-style experiments might be helpful to
clarify the robustness of the coaddition procedure.

\begin{figure}
%\plotone{sesto_fig6.ps}
\plotfiddle{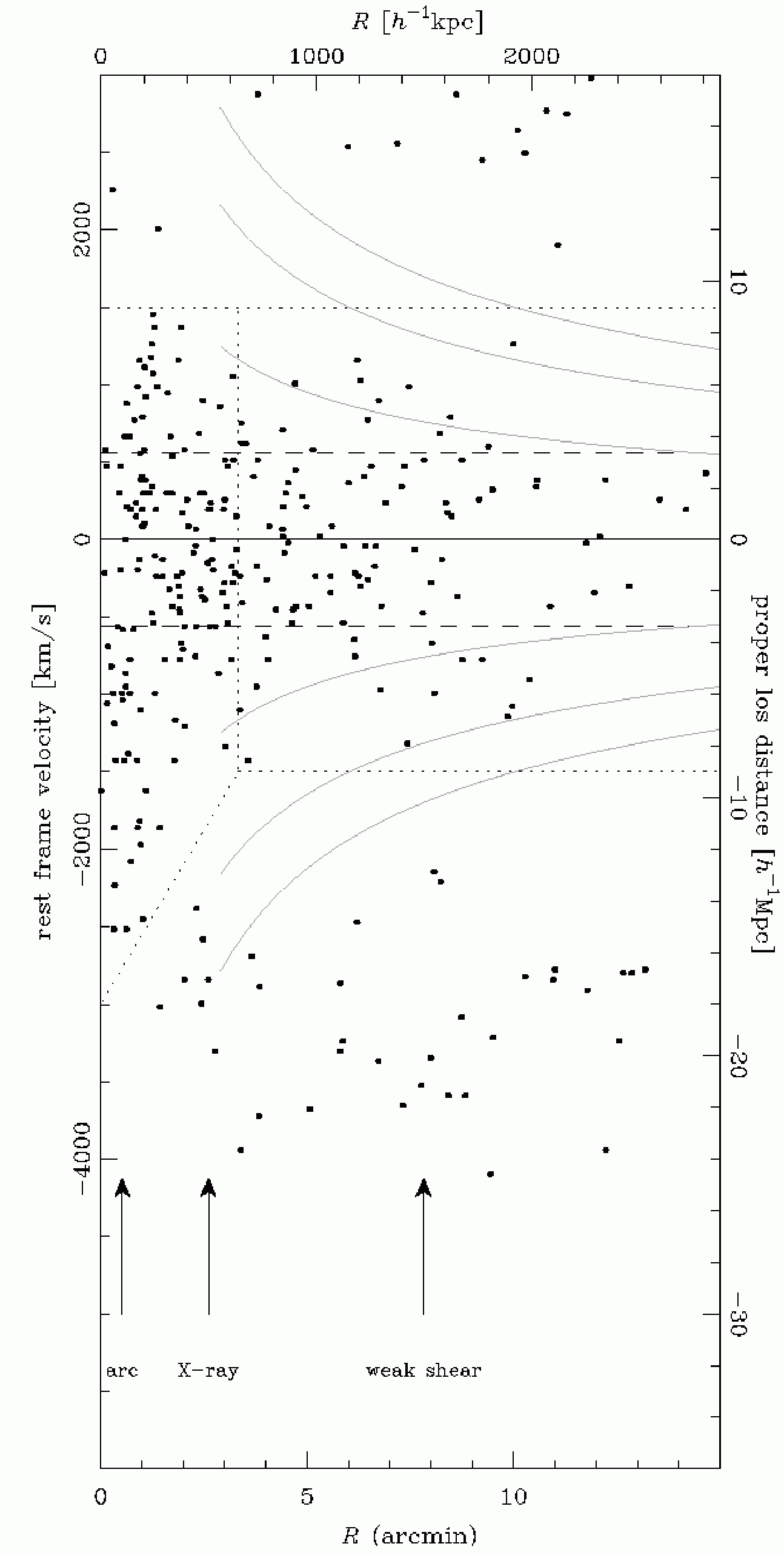}{2.2truein}{90}{28}{28}{173}{0}
\caption{Velocities versus cluster-centric radius for the rich
cluster Cl0024+16 from the CFHT survey of Czoske et al (2001). For
many years this cluster was considered to have a very high
velocity dispersion and a surprisingly low X-ray luminosity. The
authors now propose that the innermost velocity distribution
indicates a head-on collision of two systems of comparable mass.
Although not an unique explanation, the cluster serves as a
warning to those interested in cooadding velocity data over many
assumed regular clusters.}
\end{figure}

Soon after the first X-ray detections of hot gas from the
intracluster medium (ICM) were detected, it was realised that most
of the baryons in clusters are in gaseous form. Assuming an
isothermal distribution, the surface brightness profiles offer a
valuable probe of cluster mass disribution, and the X-ray
morphology indicates the evolutionary state. Clusters are expected
to continuously assemble in popular hierarchical models and we can
expect to directly witness this growth via shock heating events.

XMM and Chandra are transforming this field and this conference
has provided one of the first opportunities for us to discuss the
assumptions made in analysing earlier data. Monique Arnaud
illustrated some of this progress in her excellent review.
Although merging substructures produce shock-heating Chandra
reveals unforeseen complexities, for example in the ``cold fronts"
seen in some clusters which suggest the associated timescales may
have been underestimated. Monique demonstrated several lines of
evidence for the continuing assembly of clusters (X-ray
substructure, anisotropic accretion etc). However, even in the
well-studied case of the outskirts of the Coma cluster, Uli Briel
demonstrated significant uncertainties in understanding the merger
timescales involved. Such data are thus {\em illustrative} of
hierarchical assembly but may not give us the needed {\em
quantitative} verification of mass assembly.

The most striking aspect of the new high quality data is the
growing evidence for non-gravitational processes in understanding
the physics of the ICM. Whilst temperature profiles obtained with
ASCA and BeppoSAX were often discrepant (as discussed by De
Grandi), the XMM data gives reasonable support to the assumption
of isothermal cores. However, the expected self-similar ``scaling
laws" between the gas mass, virial radius and X-ray temperature,
e.g. in $L_x \propto T^2$ are not obeyed. The steep $L_x-T$
relation implies non-gravitational processes, almost certainly
associated with additional heating mechanisms.

Theoretical explanations for the steep $L_x-T$ relation were the
subject of almost an entire afternoon session. There was a
bewildering number of ideas. Fabio Governato reviewed for us how
feedback heating can be incorporated in numerical simulations and
showed that a steep relation cannot be easily explained via
conventional heating sources such as supernovae. Paolo Tozzi
demonstrated that, with an energy of 1-2 kev/baryon, AGN are a
promising possibility but it seems there is no obvious evidence
for heating from these sources. Peter Thomas from the VIRGO
consortium can predict the slope but only at the expense of a
strong evolution in the relationship. Uros Seljak gave us an
analytical model based on a universal gas profile where departure
from self-similarity arises from a temperature-dependent hot gas
fraction.

Perhaps that glass of wine at lunchtime was not a good idea but
this was a confusing session for me. Unlike observers who compare
ASCA/BeppoSAX/XMM temperature profiles, worrying about
sensitivity, psf and background differences, when it comes to
numerical simulations, I am puzzled that there appears to be no
necessity to inter-compare results. Surely this is a first step in
convincing anyone of the believability of a particular simulation?

Additional physical complexities in the inner cool cores of
clusters were reviewed by Andy Fabian. In the physical environment
of a dense cluster core, hot gas is able to cool radiatively in a
timescale of less than $10^{9-10}$ years and a stable flow of
cooling gas to the cluster centre is expected. Almost 50\% of all
X-ray clusters show some signs of the associated temperature
gradient implying mass inflow and deposition rates of 10-100
$M_{\odot}$ yr$^{-1}$. However, XMM spectra show no sign of the
expected Fe XX lines of radiative cooling in the 13-18 \AA\ range.
Hans B\"ohringer also discussed additional temperature diagnostics
unaffected by absorption effects. Either the gas is not cooling
radiatively or there is perhaps is a balancing effects from a
hitherto-unidentified heating mechanism. And there are serious
duty cycle issues concerned with preventing cooling beyond 2-3
kev.

As usual, Andy had no shortage of ideas for resolving the puzzle,
the most intriguing from my perspective being the hypothesis of a
starkly bimodal metallicity distribution which has the desired
effect of suppressing the amount of cooling witnessed through the
diagnostic iron lines. Like most conference participants I
suspect, I have little idea where this additional complexity in
the ICM physics is taking us. The Chandra images of the Perseus
cluster discussed by Schmidt (with such enthusiasm!) show holes in
the X-ray emission where the gas appears hotter and more metal
rich. Dynamical complexities were introduced by Ettori in the
wealth of data he presented for Abell 1795. A number of speakers
introduced magnetic fields associated with radio polarisation (in
the case of Abell 2255 discussed by Govoni), radio sources
(although as Fabian remarked the gas is not obviously hotter in
these regions) or in explaining the survival of cold clouds with
associated fronts (Arnaud).

I think we can be optimistic that there is a lot to learn which
will assist in the entire interplay between gas cooling, star
formation and enrichment. Many years ago at a conference I
attended in Cambridge, Andy claimed confidently that cooling flows
provided a key mechanism for the formation of giant elliptical
galaxies. Although cooling flows remain controversial, Andy can
still assert (as Richard Bower emphasised independently) that the
physics of gas cooling in cluster cores will tell us much about
feedback and star formation: still key ingredients in galaxy
formation.

\section{Evolution of Galaxy Populations within Dense Environments}

On the last day, Pieter van Dokkum surveyed the literature on the
role that clusters continue to play in our understanding of galaxy
formation and evolution. There is a subtle shift of emphasis
however. Rather than being ``laboratories of convenience"
\footnote{Perhaps there is a better terminology here!}, galaxies
in clusters are now being studied alongside equivalent field
populations and as likely successors to well-studied Lyman break
galaxies at $z>$2.

Most of the discussion, by van Dokkum, Rosati \& Kelson,
concentrated on the role of very distant ($z>$0.5) clusters,
largely in terms of differentiating stellar ages and the ages of
mass assembly of giant ellipticals. Our large telescopes are being
flexed to their limits to secure impressive fundamental plane data
(to $z=$1.2 in 12 hour Keck exposures!) which continues to support
the notion that old stellar populations in at least some fraction
of the data. That the stars in ellipticals may be older than the
assemblies in which they now reside is illustrated in the red
mergers seen in the well-studied X-ray luminous cluster MS1054
($z$=0.83). Pieter posed the question of how common such a system
might be and gave some new examples of red mergers in other
clusters. It is good to see some progress in separating the ages
of stars from those of the accumulated mass that makes up giant
ellipticals.

\begin{figure}
\plotone{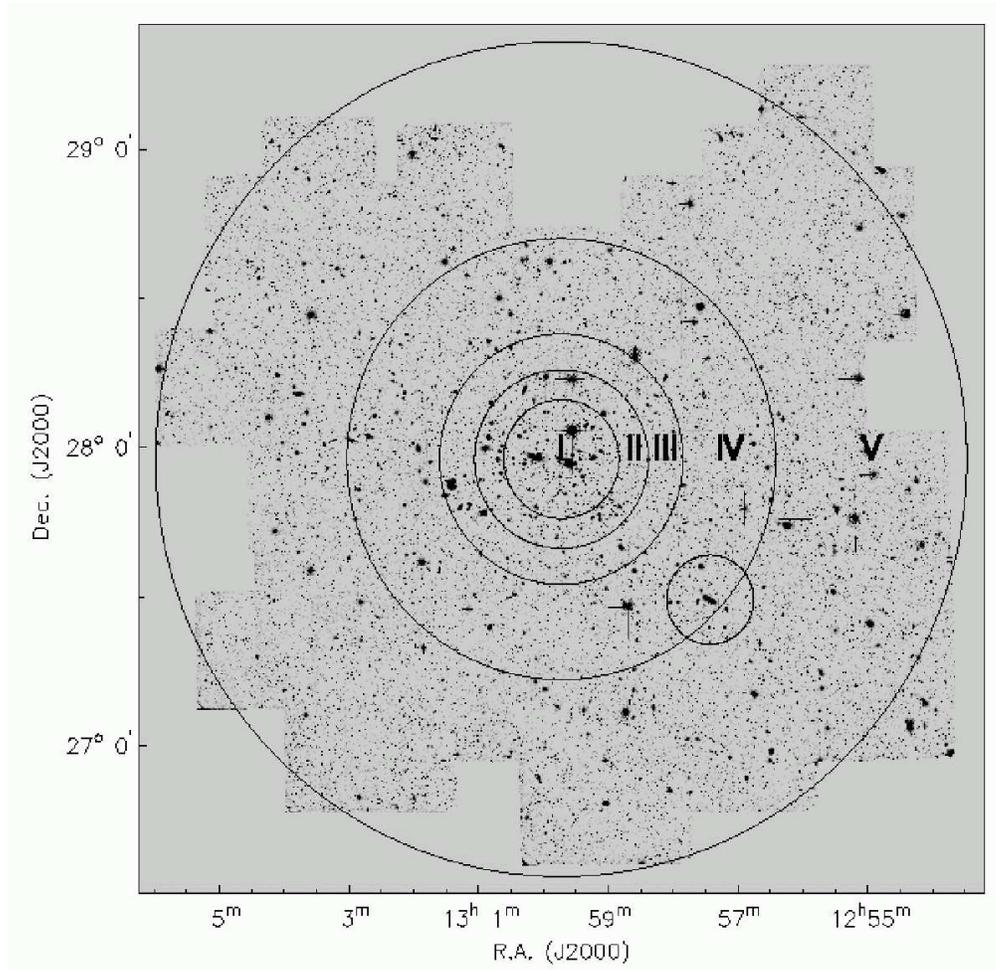}
\caption{Going back to basics: securing improved morphological and
spectroscopic data for galaxies in nearby clusters is important in
fixing the low redshift baseline for high $z$ studies. Mosaiced
CCD image of the Coma cluster from the study of Beiersbergen et al
(2001).}
\end{figure}

The biases introduced not only in selecting galaxies within
clusters (by HST morphology, by colour or infrared magnitude) but
also by how the clusters themselves were selected (X-ray,
optical..) continue to worry me. D. Fadda also reminded us how gas
rich and dusty systems may be common even in dense clusters. As
remarked earlier, it is hard to know what errors are made by
connecting data at different redshifts to delineate an
evolutionary picture. One suspects we are reliant either on
theoretical modelling (heaven forbid!) or comparative field
samples. Beyond $z>$0.5 this will be some time coming although a
number of speakers alluded to the upcoming Keck and VLT
spectroscopic surveys.

It is now 20 years since Alan Dressler published his quantitative
study of the morphology-density relation in a nearby sample of 55
rich clusters (Dressler 1980). Pieter and Bianca Poggianti urged
us to go back and improve the local samples now we have panoramic
CCD cameras and multi-object spectrographs so we can be sure of
the local fractions (Figure 7). A key issue here is the origin of
S0s; the Morphs team (Dressler et al 1997) proposed their recent
demise from gas stripping and tidal effects and presented a strong
claim for an evolving E/S0 fraction with lookback time. Pieter
urged us to be much more cautious in differentiating Es and S0s;
with resolved spectroscopy of $z>$0.5 galaxies feasible on many
telescopes, this appears to be a profitable route in conjunction
with HST data.

\section{Parting Thoughts}

This has been a fascinating meeting and we have a lot to be
thankful for. Foremost we have amazing observational facilities
capable of finding clusters and studying them to great depths in
complementary ways. We now can resolve clusters at the arcsec
level in X-rays and locate their concentrated masses with S-Z and
weak lensing techniques. We can also locate local clusters as
byproducts in the comprehensive redshift surveys being undertaken.
Instead of viewing these as competing techniques we should exploit
them all as complementary probes to check our physical
assumptions.

We also have the opportunity to learn new physics in the cluster
cores from the paradoxes emerging from the steeper $L_x-T$
relations and the absence of lines of radiative cooling in those
with temperature gradients. It should perhaps be no surprise that
simple gravitational physics is not enough to explain what we see.
The fall-out in our understanding of galaxy and cluster formation
from resolving this paradox could be a big one.

As the concluding speaker I want to thank Drs Mardirossian and
Mezzetti for all the necessary arrangements here at beautiful
Sesto Pusteria and to the Organising Committee for their hard
work. We should in particular thank Stefano Borgani for his
obvious insight in constructing a thoughtful scientific program
and amazing energy he and his colleagues invested in making this a
great meeting!

\end{document}